\documentclass[letterpaper,aps,showpacs,floatfix,10pt,prc]{revtex4-1}
\usepackage{graphicx}
\usepackage{amsmath}
\usepackage{graphicx}
\usepackage{comment}
\usepackage{float}

\begin{document}
\title{Towards a Deeper Understanding of How Experiments Constrain
the Underlying Physics of Heavy-Ion Collisions}
\author{Evan Sangaline}
\affiliation{Department of Physics and Astronomy and National Superconducting Cyclotron Laboratory\\
Michigan State University, East Lansing, MI 48824~~USA}
\author{Scott Pratt}
\affiliation{Department of Physics and Astronomy and National Superconducting Cyclotron Laboratory\\
Michigan State University, East Lansing, MI 48824~~USA}
\date{\today}


\begin{abstract}
Recent work has provided the means to rigorously determine properties of super-hadronic matter from experimental data through the application of broad scale modeling of high-energy nuclear collisions within a Bayesian framework. These studies have provided unprecedented statistical inferences about the physics underlying nuclear collisions by virtue of simultaneously considering a wide range of model parameters and experimental observables. Notably, this approach has been used to constrain both the QCD equation of state and the shear viscosity above the quark-hadron transition. Although the inferences themselves have a clear meaning, the complex nature of the relationships between model parameters and observables have remained relatively obscure. We present here a novel extension of the standard Bayesian Markov Chain Monte Carlo approach that allows for the quantitative determination of how inferences of model parameters are driven by experimental measurements and their uncertainties. This technique is then applied in the context of heavy ion collisions in order to explore previous results in greater depth. The resulting relationships are useful for identifying model weaknesses, prioritizing future experimental measurements, and most importantly: developing an intuition for the role that different observables play in constraining our understanding of the underlying physics.
\end{abstract}

\maketitle

\section{Introduction}
\label{sec:intro}

Relativistic nuclear collisions provide a unique opportunity to create and study matter at temperatures and energy densities well above the boundary where hadronic degrees of freedom become irrelevant and are replaced with quark and gluon degrees of freedom. Whereas the energy density inside a hadron might be on the order of 100 MeV/fm$^3$, high-energy nuclear collisions produce mesoscopic regions where the average energy density can exceed 10 GeV/fm$^3$. Unfortunately, they are far from ideal environments for extracting the properties of super-hadronic matter. Only the asymptotic momenta of outgoing particles are experimentally accessible, so three-dimensional models of the dynamics are essential for interpreting results and forming rigorous conclusions about either the nature and properties of the matter or about the evolution itself. One serious limitation is that the exact nature of the initial deposition and dispersion of energy, momentum, charge, and baryon number in heavy-ion collisions is currently only partially understood. Nearly any experimental measurement will be sensitive to these initial conditions as well as to combined contributions from the proceeding stages of rapid expansion and cooling through partially equilibrated QGP and hadronic phases. This entangles contributions from the various components of the underlying physics and obscures the interpretation of many experimental results. 

The development of models that can reliably mimic the various complexities of heavy-ion collisions has been a key focus of the field and a necessary step towards condensing experimental results into quantitative statements about the underlying physics. Unfortunately, determining physical quantities like the shear and bulk viscosities of QGP is non-trivial even with high quality models in place. Experimental data can be largely described using these models but the conclusions drawn depend strongly on the validity of the initial state model and other model assumptions. This problem is only confounded by the computational complexity of typical heavy-ion collision models which, in many cases, makes a full exploration of model parameter space prohibitively expensive and, in turn, encourages simplifying assumptions.

The situation of dealing with very complicated models that map high dimensional parameter spaces to rich heterogenous observational data sets is a common one. It's faced in disparate fields of science such as biology, econometrics, and cosmology. The common solution to this problem is to apply a Bayesian inference approach where a posterior distribution over model parameters can be mapped out based on their consistency with a set of observables using a Markov Chain Monte Carlo (MCMC) or a related approach \cite{Habib:2007ca,Gomez:2012ak}. This allows for quantitative conclusions about the underlying parameters to be made given the validity of the model and prior assumptions. 

A primary difficulty that has historically limited the applicability of a Bayesian approach in heavy-ion physics has been the large number of model evaluations necessary to accurately map out a high dimensional posterior distribution. This, coupled with the already computationally expensive nature of the simulations, makes a direct application of an MCMC approach essentially impossible with currently available hardware. Recent progress in methodology has overcome this difficulty through the use of Gaussian process emulators which can be trained to accurately reproduce the results of actual simulated models. The reduced computational load has opened the door to quantitative analyses of model parameters which has allowed for systematic constraints of the nuclear equation of state and of the QGP shear viscosity \cite{Novak:2013bqa,Pratt:2015zsa}.

Stating the single point in the multidimensional parameter space that maximizes the likelihood disregards the uncertainty with which the number is stated. In contrast, an MCMC exploration of parameter space provides not only the uncertainties for each parameter, but the covariances with other parameters along with the complete shape of the posterior likelihood. The methods of \cite{Novak:2013bqa} indeed provide the complete likelihood distribution, and are now extracting the field's first rigorous quantitative conclusions from heavy-ion collisions at the highest energies.

Although the Bayesian approach has proved fruitful in nuclear physics, progress in understanding the complexities of generated posterior distributions and how they're driven by the experimental data has been slow. It is relatively straightforward to understand how parameters drive observables through varying them and seeing what changes. Going in the opposite direction, seeing how the measurements drive the inferences about parameters, is more challenging for several reasons. For example, changing a certain parameter might readily alter a given observable, but there may exist some combination of other parameters that can compensate without affecting the remaining observables. Even an observable which does not seem to directly depend on a given parameter, might help constrain other parameters, which then helps constrain the first parameter. In deciding which observables to either measure or to better measure, one would like to know how the width of the posterior likelihood distribution is affected by reducing the uncertainty of a given observable. These relations are critical to gaining insight into understanding not only the degree to which model parameters are being constrained, but how and why.  Here, we present a new set of techniques for addressing all of these questions and then apply the techniques to the 14-parameter analysis first presented in \cite{Pratt:2015zsa}. This analysis represents the field's first quantitative evaluation of much of the field's consensus-based understanding about which observables are truly responsible for addressing the community's most basic questions about the bulk properties of nuclear matter and about the evolution of a high-energy heavy-ion collision.

In the next two sections we review the model, data and techniques used to generate our previous results in \cite{Novak:2013bqa,Pratt:2015zsa}, presenting a wider range of results from the projections of the MCMC procedure than were presented in \cite{Pratt:2015zsa}. In Sec. \ref{sec:sensitivity} we present techniques for determining linear relations between observables and parameters and between uncertainties in observables and widths of the posterior distribution in parameter space. We apply these techniques to the heavy-ion analysis described in Sec.s \ref{sec:model} and \ref{sec:mcmc}, and focus on the ramifications for extracting the equation of state and viscosity. Results are summarized and an outlook is presented in the final section. 

\section{Model and Data Overview}
\label{sec:model}

The details of the model are expounded in more detail in \cite{Novak:2013bqa,Pratt:2015zsa}, but we will briefly review the basics and focus on describing the 14 parameters varied in this analysis along with the set of observables. Our model consists of the generation of an initial state which is then fed into a 2-dimensional hydrodynamics simulation followed by transition to a microscopic simulation, known as a hadronic cascade, at a transition temperature of $T_{0}=165\ \text{MeV}$. An assumed symmetry based on an invariance to boosts along the beam axis makes it possible to approximate three-dimensional treatment with a two-dimensional model that discretizes the two transverse coordinates. 

Ten model parameters are used to vary the initial state at a time $\tau_0=0.8$ fm/$c$, two are used to determine the equation of state in the hydrodynamic stage, and the last two are used to determine the shear viscosity and it's temperature dependence in the hydrodynamic stage. The shear viscosity is parameterized as 
\begin{equation}
\frac{\eta}{s}
=\left(\frac{\eta}{s}\right)_{0}+\eta'\ln\frac{T}{T_0}
\end{equation}
where $(\eta/s)_0$ is the viscosity at $T_0$ and $\eta'$ describes the temperature dependence. These two parameters will be featured in this paper due to their relatively intuitive meaning and relationship with common observables. Two parameters also encapsulate the equation of state. The speed of sound, $c_s$, as a function of the energy density, $\epsilon$, is described as
\begin{eqnarray}
c_s^2(\epsilon)&=&c_s^2(\epsilon_h)+\left(\frac{1}{3}-c_s^2(\epsilon_h)\right)
\frac{X_0x+x^2}{X_0x+x^2+X'^2},\\
\nonumber
X_0&=&X'Rc_s(\epsilon)\sqrt{12},~~x\equiv \ln(\epsilon/\epsilon_h),
\end{eqnarray}
where $\epsilon_h$ and $c_s(\epsilon_h)$ are the energy density and speed of sound of a hadron gas at the transition temperature, $T_0=165$ MeV. These quantities are calculated by considering a gas of non-interacting hadrons using the masses and spins of particles from the Particle Data Group \cite{pdg}. All resonances with masses below 2 GeV/$c^2$ were included. With this prescription, the equation of state is continuous at $T_0$. The parameter $R$ describe the behavior of $c_s$ just above $T_0$ and the parameter $X'$ provides a scale at which the speed of sound approaches $1/3$. Increasing $X'$ lowers the speed of sound mainly at high energy density or temperature, and increasing $R$ increases the speed of sound, mainly just above $T_0$. 

Five of the ten initial state parameters apply only to the description of RHIC (Relativistic Heavy Ion Collider at Brookhaven National Laboratory) data. Only data from Au+Au collisions at $100A$ GeV + $100A$ GeV is considered. The other five described the initial state for Pb+Pb collisions at $1.38A$ TeV + $1.38A$ TeV from the LHC (Large Hadron Collider at CERN). The initial transverse energy density profile that instantiates the hydrodynamics has the form \cite{Novak:2013bqa}
\begin{eqnarray}
\epsilon(x,y)&=&f_{\rm wn}\epsilon_{\rm wn}(x,y) + (1-f_{\rm wn})\epsilon_{\rm sat}(x,y),
\end{eqnarray}
where the parameter $f_{\rm wn}$ describes the weighting between two other parameterized forms, $\epsilon_{\rm wn}$ which is the wounded nucleon form \cite{Miller:2007ri}, and $\epsilon_{\rm sat}$, which describes a form more in line with some ideas of saturation, similarly to \cite{Drescher:2007cd}. 
\begin{eqnarray}
\label{eq:wn}
\epsilon_{\rm wn}(x,y)&=& 
Z_\epsilon\frac{(dE_\perp/dy)_{pp}\sigma_{\rm nn}}{2\sigma_{\rm sat.}}
T_A(x,y)\left(1-\exp(-T_B(x,y)\sigma_{\rm sat})\right),\\
\nonumber
&&+Z_\epsilon\frac{(dE/dy)_{pp}\sigma_{\rm nn}}{2\sigma_{\rm sat.}}T_B(x,y)\left(1-\exp(-T_A(x,y)\sigma_{\rm sat})\right),\\
\label{eq:sat}
\epsilon_{\rm sat}(x,y)&=&Z_\epsilon\frac{(dE/dy)_{pp}\sigma_{\rm nn}}{\sigma_{\rm sat.}}
T_{\rm min}(x,y)\left(1-\exp(-T_{\rm max}(x,y)\sigma_{\rm sat})\right),\\
\nonumber
T_{\rm min}&\equiv&\frac{2T_AT_B}{T_A+T_B},~~T_{\rm max}\equiv(T_A+T_B)/2.
\end{eqnarray}
Here $T_A$ and $T_B$ are the areal densities, the projections of the baryon density of a Au or Pb nucleus onto the transverse plane, and have dimensions of number per area. In the wounded nucleon model, each nucleon that comes within the nucleon-nucleon cross section, $\sigma_{nn}=42$ mb at RHIC energies and 73 mb at LHC energies, of one of the other nucleons, known as a participant, contributes to the energy density. If the parameter $\sigma_{\rm sat}$ is set equal to the nucleon-nucleon cross section then each nucleon can only contribute once and additional collisions do not increase the energy density. Relaxing $\sigma_{\rm sat}<\sigma_{nn}$ allows the particle to contribute multiple times. As $\sigma_{\rm sat}$ approaches zero the form gives binary scaling and the resulting energy density is proportional to $T_AT_B$. In the saturated form, the energy density is principally determined by the smaller of the two areal densities. Thus, if one nucleon overlaps 5 other nucleons, the energy density will be only slightly less than if it overlapped 10.

The parameter $Z_\epsilon$ represents the energy per unit rapidity per nucleon collision in a dilute reaction relative to the measurement of a $pp$ collision, and both forms satisfy the constraint that for a diffuse overlap of nucleons, i.e. $T_A,T_B\ll \sigma_{pp}$ the energy density scales as binary collisions, $\epsilon(x,y)\rightarrow \sigma_{\rm pp}(dE/dy)_{\rm pp}T_AT_B$ with $Z_\epsilon\approx 1$. However, because this is the energy density at $\tau_0=0.8$ fm/$c$, and not the final energy, and because the energy density includes longitudinal motion, it differs from the measured transverse energy,  and $Z_\epsilon$ was allowed to vary over a small range near unity. For our purposes $(dE/dy)_{\rm pp}$ was set to 2.69 and 6.0 GeV for the two beam energies, though this number is somewhat arbitrary because it is multiplied by $Z_\epsilon$. Finally, the fifth parameter describes the initial transverse flow, which is assumed to have the form
\begin{equation}
\label{eq:univflow}
\frac{T_{0i}}{T_{00}}=F_0\frac{- \partial_i T_{00}}{2T_{00}} \tau,
\end{equation}
where $T_{\alpha\beta}$ is the stress-energy tensor. For a traceless stress energy tensor, which would be expected for non-interacting gluon fields or collision-less massless particles or for conformal hydrodynamics, and also assuming boost invariance, $F_0$ would go to unity \cite{Vredevoogd:2008id}. The final parameter adjusted the initial anisotropy of the stress-energy tensor,
\begin{equation}
T_{xx}=T_{yy}=P(1+2\tau'_{xx}), ~~~T_{zz}=P(1-\tau'_{xx}).
\end{equation}
This form is traceless $(1/3)\sum_{i=1,2,3}T_{ii}=P$ and describes the initial shear, which in the Israel Stewart form of hydrodynamics is a dynamical variable which relaxes toward the Navier Stokes value. Parameters are summarized in Table \ref{table:parameters}. 

\begin{table}
\begin{tabular}{|c|c|l|}\hline
Parameter & Min, Max (RHIC/LHC values)& Description\\ \hline
$R$ & -0.9, 2.0 & Equation of State \\
$X'$ & 0.5, 5.0 & Equation of State \\ 
$(\eta/s)_0$ & 0.02, 0.5 & Viscosity at $T_0$\\ 
$\eta'$ & 0.0, 3.0 & Temperature dependence of viscosity\\ 
$Z_\epsilon$ & 0.8 / 0.7, 1.25 / 1.4 & Energy normalization ratio\\
$\sigma_{\rm sat}$(mb) & 22 / 38, 44 / 76 & Saturation cross section\\
$f_{\rm wn}$ & 0.0 / 0.0, 1.0 / 1.0 & Weight of wounded-nucleon parameterization\\
$F_0$ & 0.2 / 0.2, 1.0 / 1.0 & Initial transverse flow\\
$\tau'_{xx}$ & 0.0 / 0.0, 1.0 / 1.0 & Initial anisotropy of stress-energy tensor\\ \hline
\end{tabular}
\caption{\label{table:parameters}
Fourteen model parameters}
\end{table}

The model is used to produce a set of 30 observables that correspond to experimental measurements performed at RHIC and the LHC. These encompass distilled information describing spectra, elliptic flow, and femtoscopic correlations in central and mid-central collisions. These observables were chosen because they all have been shown to characterize features of thermalized bulk matter. Fifteen of the observables correspond to Au+Au collisions at RHIC while the other fifteen correspond to Pb+Pb collisions at the LHC. Observables were taken from two centrality classes, the top $0-5\%$ centrality cut and the set of collisions in the $20-30\%$ centrality cut. It was felt that additional data with centrality between these values would be redundant because of the smooth behavior of the observations with centrality. Further,  it would be difficult to assign uncertainties for observables from more peripheral collisions due to the difficulty in justifying hydrodynamic treatments for systems whose size is not larger than a thermal wavelength, and where a significant fraction of the transverse profile is in the corona, which does not fully thermalize.

Three classes of measurements were considered. The first and most basic is spectra and yields. These were distilled to four numbers for each centrality. Three were the mean transverse momenta, $p_t$, for pions, kaons and protons. The average was taken over a finite range of $p_t$, limited at the low end by experimental constraints and cut off at the high end to minimize the effects of jets which are non-thermal features outside of this analysis. The fourth measurement was the multiplicity of pions within the same $p_t$ range. Because chemical equilibrium was assumed, rather than parameterized, and because baryon annihilation was not included, the yields of kaons and protons were not included. In \cite{Novak:2013bqa} it was shown that model runs that had the same $\langle p_t\rangle$ had the same spectral shapes, so no resolving power was lost by considering only one number to characterize the spectra. Those spectra with the same $\langle p_t\rangle$ as the data also provided good descriptions of the experimental spectral shapes \cite{Novak:2013bqa,Pratt:2015zsa}. 

The second class of observables is comprised of femtoscopic radii extracted from two-pion correlations at small relative momentum. Extracting these radii has long been a staple of the field \cite{Lisa:2005dd}. The gaussian radii are functions of momentum, and describe the size and shape of the outgoing phase space cloud of the given momentum. The three radii, $R_{\rm out}, R_{\rm side}$ and $R_{\rm long}$, describe the transverse dimension parallel to the momentum, the transverse size perpendicular to the momentum and the size along the beam axis respectively. For this analysis, the radii were averaged over the various bins in transverse momentum, so that three observables encapsulated the data for each centrality class.

The final observable was $v_2$, which characterizes the anisotropic transverse flow which is driven by the elliptical shape of the initial transverse profile. The anisotropy, $v_2\equiv\langle \cos 2\phi\rangle$ in its simplest definition, was only considered in the 20-30\% centrality bin. This bin was chosen because the model used smooth profiles, generated from the average aerial profiles for a given impact parameter, and neglected the lumpy conditions which one would expect from the finite number of scatterers in a Au or Pb nucleus. Even though this bin is probably the least affected by lumps, the experimental value was reduced by a factor of 0.91 to more fairly compare to the model \cite{Pratt:2015zsa}. The next major analysis planned by this group will incorporate fluctuating initial conditions, and in addition to more accurately calculating $v_2$, could also address the higher components $v_3, v_4\cdots$, which are purely driven by the fluctuations. Because $v_2$ rises nearly linearly with transverse momentum, the value of $v_2$ was averaged over $p_t$, with the values weighted by $p_t$, so that bins with higher $p_t$ were relatively more important. The linear weighting was motivated by performing a principal component analysis of the binned values to find the combination that best captured the variability in the model runs. Table \ref{table:pcaobservables} summarizes the 30 observables in this analysis.

\begin{table}
\begin{tabular}{|c|c|c|c|c|c|c|}\hline
observable & exp. value & $p_t$ weighting & centrality & collaboration\\ \hline
$v_{2,\pi^+\pi^-}$ & 8.14\%& ave. over 11 $p_t$ bins from 160 MeV/$c$ to 1 GeV/$c$ & 20-30\% & STAR \cite{Adams:2004bi}\\
$R_{\rm out}$ & 5.28 fm &ave. over 4 $p_t$ bins from 150-500 MeV/$c$ & 0-5\% & STAR \cite{Abelev:2009tp}\\
$R_{\rm side}$ & 4.81 fm & ave. over 4 $p_t$ bins from 150-500 MeV/$c$& 0-5\% & STAR \cite{Abelev:2009tp} \\
$R_{\rm long}$ & 5.47 fm & ave. over 4 $p_t$ bins from 150-500 MeV/$c$& 0-5\% & STAR \cite{Abelev:2009tp} \\
$R_{\rm out}$ & 4.27 fm & ave. over 4 $p_t$ bins from 150-500 MeV/$c$& 20-30\% & STAR \cite{Abelev:2009tp} \\
$R_{\rm side}$ & 3.99 fm &  ave. over 4 $p_t$ bins from 150-500 MeV/$c$& 20-30\% & STAR \cite{Abelev:2009tp} \\
$R_{\rm long}$ & 4.53 fm & ave. over 4 $p_t$ bins from 150-500 MeV/$c$& 20-30\% & STAR \cite{Abelev:2009tp} \\
$\langle p_t\rangle_{\pi^+\pi^-}$ & 494.4 MeV & 0.2 GeV/$c < p_t< 1.2$ GeV/$c$ & 0-5\% & PHENIX \cite{Adler:2003cb} \\
$\langle p_t\rangle_{K^+K^-}$ & 796 MeV & 0.4 GeV/$c < p_t< 1.6$ GeV/$c$ & 0-5\% & PHENIX \cite{Adler:2003cb} \\
$\langle p_t\rangle_{p\bar{p}}$ & 1.135 GeV & 0.6 GeV/$c < p_t< 2.0$ GeV/$c$ & 0-5\% & PHENIX \cite{Adler:2003cb} \\
$\langle p_t\rangle_{\pi^+\pi^-}$ & 487.5 MeV & 0.2 GeV/$c < p_t< 1.2$ GeV/$c$ & 20-30\% & PHENIX \cite{Adler:2003cb} \\
$\langle p_t\rangle_{K^+K^-}$ & 792 MeV &0.4 GeV/$c < p_t< 1.6$ GeV/$c$ & 20-30\% & PHENIX \cite{Adler:2003cb} \\
$\langle p_t\rangle_{p\bar{p}}$ & 1.111 GeV & 0.6 GeV/$c < p_t< 2.0$ GeV/$c$ & 20-30\% & PHENIX \cite{Adler:2003cb} \\
$\pi^+\pi^-$ yield & 422 & 0.2 GeV/$c < p_t< 1.2$ GeV/$c$ & 0-5\% & PHENIX \cite{Adler:2003cb}\\
$\pi^+\pi^-$ yield & 188.7 & 0.2 GeV/$c < p_t< 1.2$ GeV/$c$ & 20-30\% & PHENIX \cite{Adler:2003cb}\\
\hline
$v_{2,\pi^+\pi^-}$ & 9.56\% & ave. over 11 $p_t$ bins from 0.15 to 1.2 GeV/$c$ & 20-30\% & ALICE \cite{Abelev:2014pua}\\
$R_{\rm out}$ & 5.46 fm & ave. over 7 $p_t$ bins from 200 to 900 MeV/$c$ & 0-5\% & ALICE \cite{Graczykowski:2014hoa} \\
$R_{\rm side}$ & 5.32 fm & ave. over 7 $p_t$ bins from 200 to 900 MeV/$c$  & 0-5\% & ALICE \cite{Graczykowski:2014hoa}  \\
$R_{\rm long}$ & 5.72 fm & ave. over 7 $p_t$ bins from 200 to 900 MeV/$c$  & 0-5\% & ALICE \cite{Graczykowski:2014hoa}  \\
$R_{\rm out}$ & 4.07 fm & ave. over 7 $p_t$ bins from 200 to 900 MeV/$c$  & 20-30\% & ALICE \cite{Graczykowski:2014hoa}  \\
$R_{\rm side}$ & 4.12 fm & ave. over 7 $p_t$ bins from 200 to 900 MeV/$c$  & 20-30\% & ALICE \cite{Graczykowski:2014hoa}  \\
$R_{\rm long}$ & 4.41 fm & ave. over 7 $p_t$ bins from 200 to 900 MeV/$c$  & 20-30\% & ALICE \cite{Abelev:2009tp} \\
$\langle p_t\rangle_{\pi^+\pi^-}$ & 459.1 MeV & 0.1 GeV/$c < p_t< 1.2$ GeV/$c$ & 0-5\% & ALICE \cite{Abelev:2013vea} \\
$\langle p_t\rangle_{K^+K^-}$ & 775 MeV & 0.2 GeV/$c < p_t< 1.6$ GeV/$c$ & 0-5\% & ALICE \cite{Abelev:2013vea} \\
$\langle p_t\rangle_{p\bar{p}}$ & 1.137 GeV & 0.2 GeV/$c < p_t< 2.0$ GeV/$c$ & 0-5\% & ALICE \cite{Abelev:2013vea} \\
$\langle p_t\rangle_{\pi^+\pi^-}$ & 455 MeV & 0.1 GeV/$c < p_t< 1.2$ GeV/$c$ & 20-30\% & ALICE \cite{Abelev:2013vea} \\
$\langle p_t\rangle_{K^+K^-}$ & 758 MeV & 0.2 GeV/$c < p_t< 1.6$ GeV/$c$ & 20-30\% & ALICE \cite{Abelev:2013vea} \\
$\langle p_t\rangle_{p\bar{p}}$ & 1.110 MeV & 0.2 GeV/$c < p_t< 2.0$ GeV/$c$ & 20-30\% & ALICE \cite{Abelev:2013vea} \\
$\pi^+\pi^-$ yield & 1258 & 0.2 GeV/$c < p_t< 1.2$ GeV/$c$ & 0-5\% & ALICE \cite{Abelev:2013vea}\\
$\pi^+\pi^-$ yield & 523 & 0.2 GeV/$c < p_t< 1.2$ GeV/$c$ & 20-30\% & ALICE \cite{Abelev:2013vea}\\ \hline
\end{tabular}
\caption{\label{table:pcaobservables} Observables used to compare models to data from RHIC (STAR and PHENIX collaborations) and from ALICE (at the LHC). To account for non-flow correlations, the value of $v_2$ was reduced by 9\% from the experimental values to account for the approximation of smooth initial conditions.}
\end{table}

\section{Markov Chain Monte Carlo Procedure and Results}
\label{sec:mcmc}

The standard approach to determining the likely regions of parameters, $\vec{x}$, from comparing model values, $\vec{y}_M(\vec{x})$, to experimental values, $\vec{y}_{\rm exp}$, is through MCMC, which provides a sampling of parameters $\vec{x}$ that are chosen weighted proportional to the likelihood. For this study the likelihood is chosen to have a simple Gaussian form, 
\begin{equation}
{\mathcal L}(\vec{x})\sim \exp\left\{-\sum_a\frac{(y_{M,a}(\vec{x})-y_{{\rm exp},a})^2}{2\sigma_a^2}\right\}.
\end{equation}
The uncertainties incorporate both experimental uncertainties and the shortcomings of the model. For instance, if the equation of state and viscosity were perfectly described, and if the initial state was parameterized most correctly, the model uncertainty describes how accurately one would expect to reproduce the final-state observables given the missing physics in the model and the uncertainties in the measurements. In this study, which uses the same model output as in \cite{Pratt:2015zsa}, the uncertainties $\sigma_a$ were all set to 6\% of the experimental value. A more accurate determination of the uncertainty would require detailed model studies to estimate the impact of missing physics, and a more detailed understanding of experimental uncertainties. It would not be surprising to find that a few of these observables are understood with slightly better uncertainty or that some are somewhat more uncertain. Such an improvement should involve input from both the experimental and modeling communities. In \cite{Novak:2013bqa} the analysis was repeated with 9\% uncertainties and the widths of the resulting posterior distributions only increased by $\approx 20\%$. 

Due to the infeasibility of performing millions of full model runs, the MCMC procedure implemented an emulator in the place of the full model \cite{Novak:2013bqa,Gomez:2012ak}. The emulator uses an interpolation algorithm to determine the observables from 1200 full-model runs. The first 1000 parameter sets for the full-model runs uniformly covered the 14-dimensional space according to latin hyper-cube sampling. The last 200 parameter sets were chosen to be consistent with the likelihood as calculated from the earlier runs and to provide better coverage in the likely region. To improve efficiency, the emulator calculates principal components rather than each observable. In this way, those linear combinations of observables that stay constant throughout the model runs can be neglected. Thus, rather than evaluating all 30 observables, only 14 principal components were analyzed for this study. Even that number could be reduced because the last several components only varied by a few percent of one value of the experimental uncertainty.

\begin{figure}
\centerline{\includegraphics[width=\textwidth]{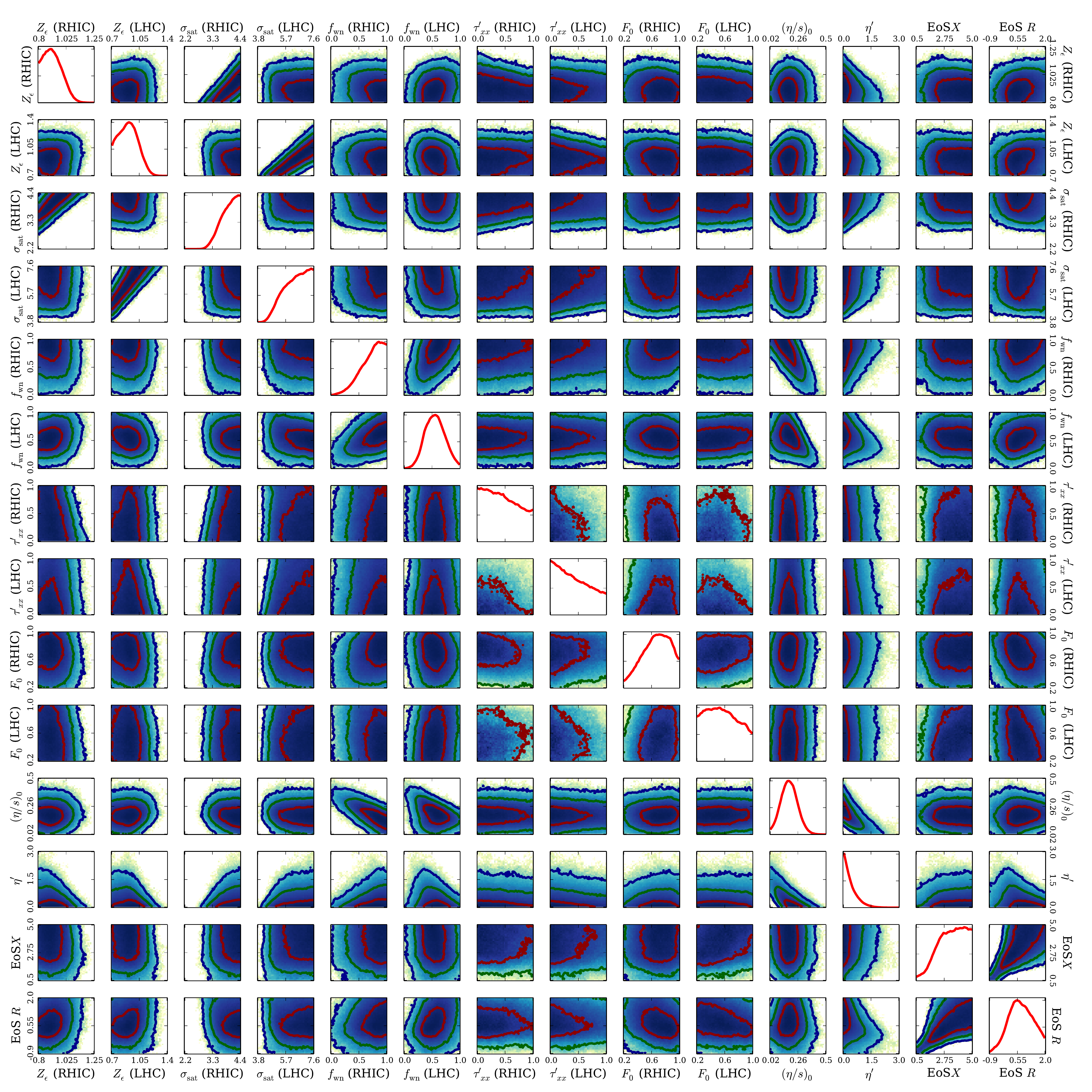}}
\caption{\label{fig:ll}
One- and two-dimensional posterior likelihood projections from the MCMC procedure. The one-dimensional projections (along the diagonal) illustrate the degree to which an individual parameter is constrained by comparing to data, whereas the off-diagonal elements illustrate how some linear combinations of two parameters might be either poorly or well constrained. The colored boundaries delineate one-, two- and three-sigma regions, where $n$-sigma refers to likelihoods of $e^{-n^2/2}$ of the maximum likelihood.}
\end{figure}
The MCMC procedure was performed with several million random steps according to the Metropolis algorithm. This provides a sampling of the posterior distribution displayed in Fig. \ref{fig:ll}. In addition to projections onto one dimension of the parameter space, the off-diagonal plots show two-dimensional projections. When the elliptic shapes of the two-dimensional projections lie at an angle, it shows that a specific linear combination of parameters may be well constrained, whereas the orthogonal combination may be poorly constrained. For example, the region of high likelihood for the projection onto the plane of two equation of state parameters, one can see that if one increases $R$ and $X'$ by similar percentages that the likelihood changes little.  Similarly, for the two viscosity parameters one finds that one can find a good match with $(\eta/s)_0\approx 0.2$ with a modest temperature dependence, $\eta'$. One can also match with lower values of $(\eta/s)_0$ if the viscosity then rises with temperature. Another instance where the posterior is off-diagonal is in the projection of $(\eta/s)_0$ and $f_{\rm wn}$. The resulting covariance corroborates the arguments that were put forward in \cite{Drescher:2007cd}.

\section{Sensitivity Studies}
\label{sec:sensitivity}

\begin{figure}
\centerline{
\includegraphics[width=0.16667\textwidth]{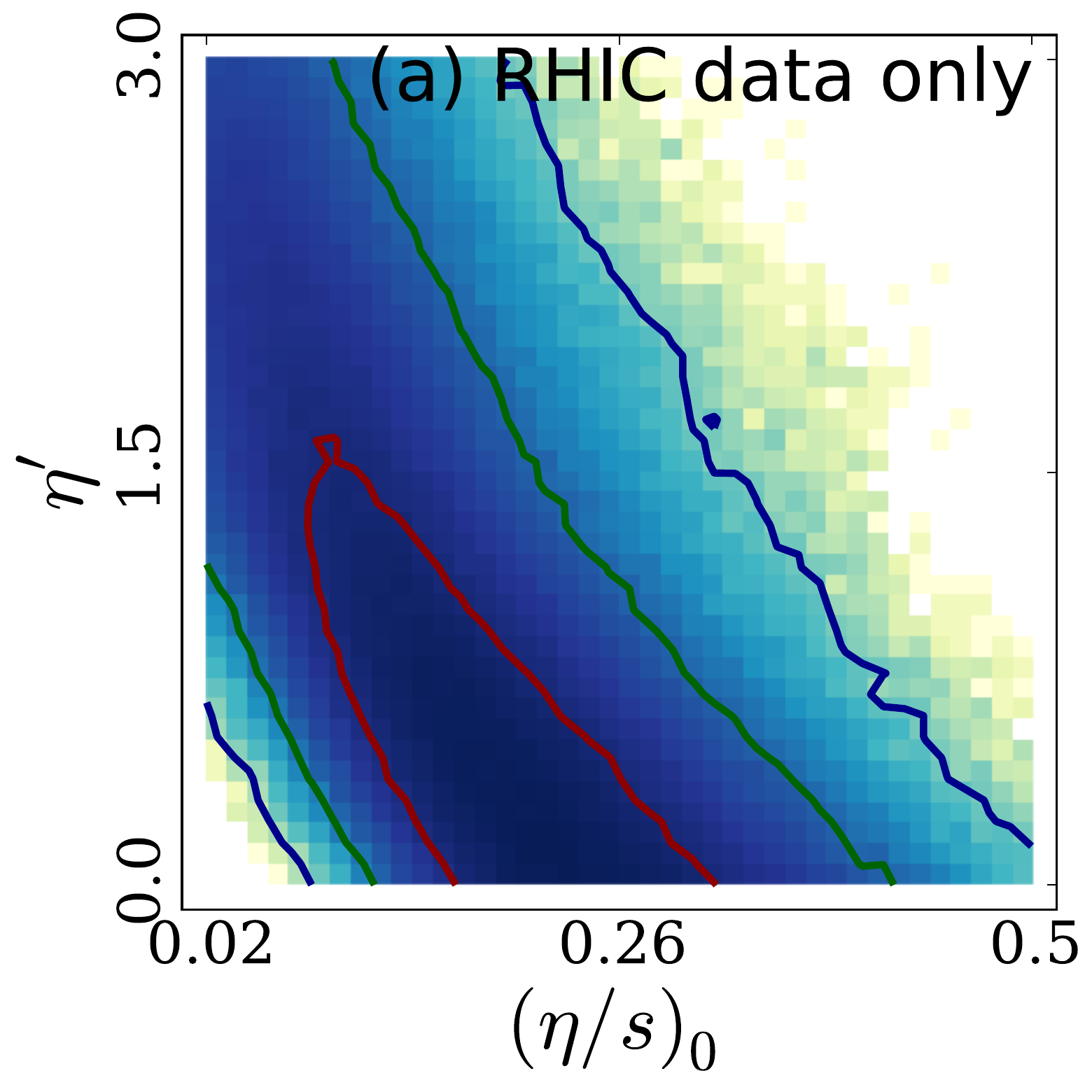}
\includegraphics[width=0.16667\textwidth]{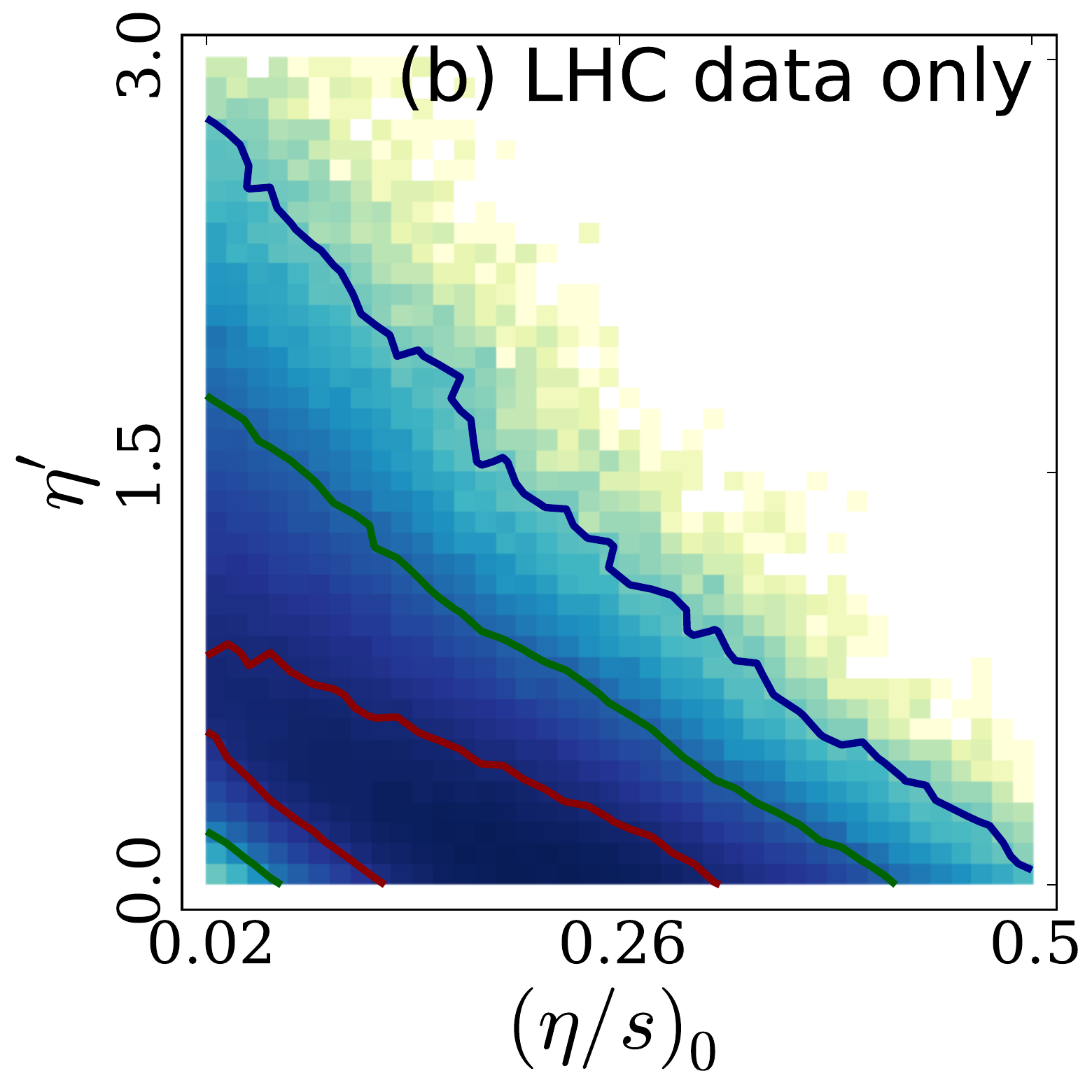}
\includegraphics[width=0.16667\textwidth]{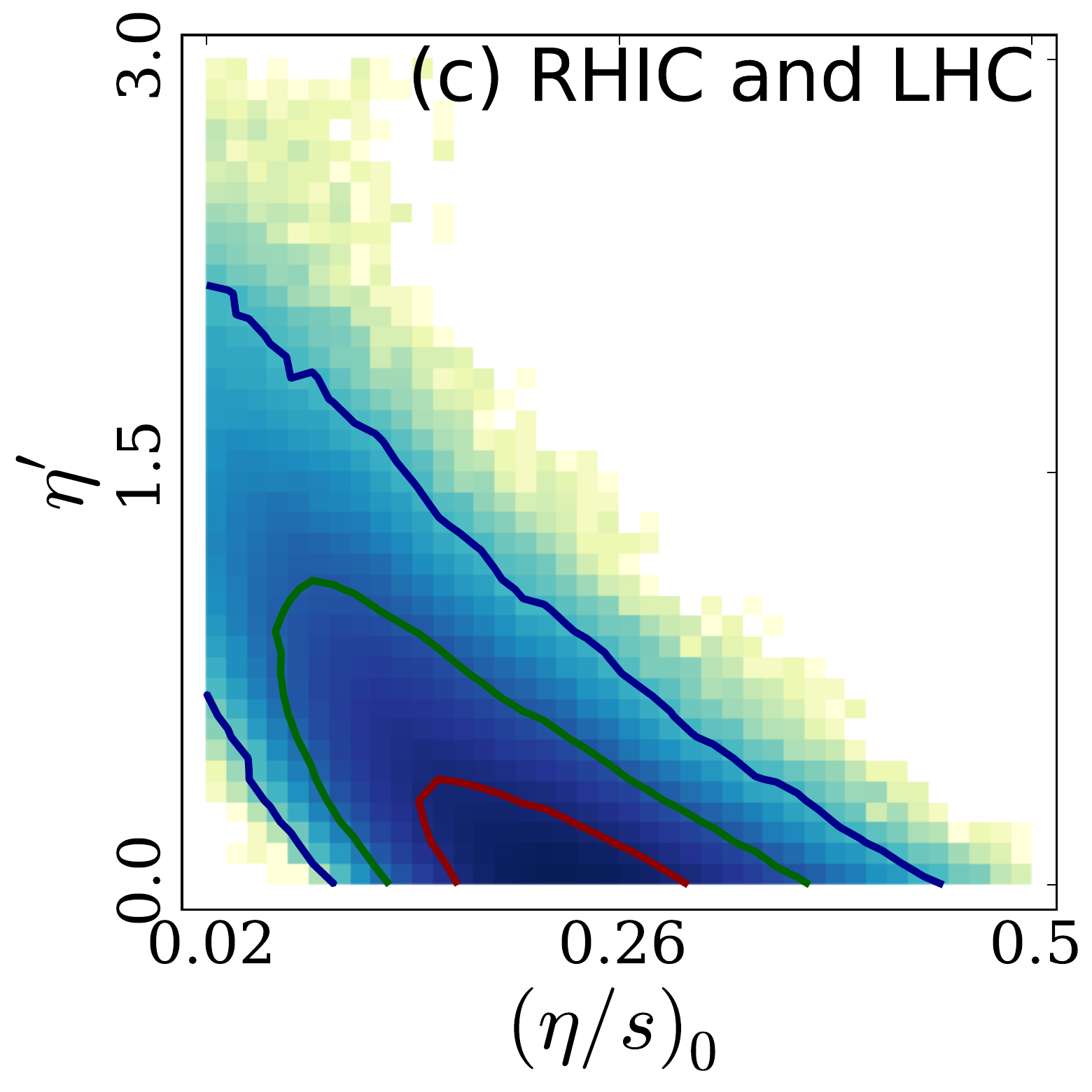}
\includegraphics[width=0.166667\textwidth]{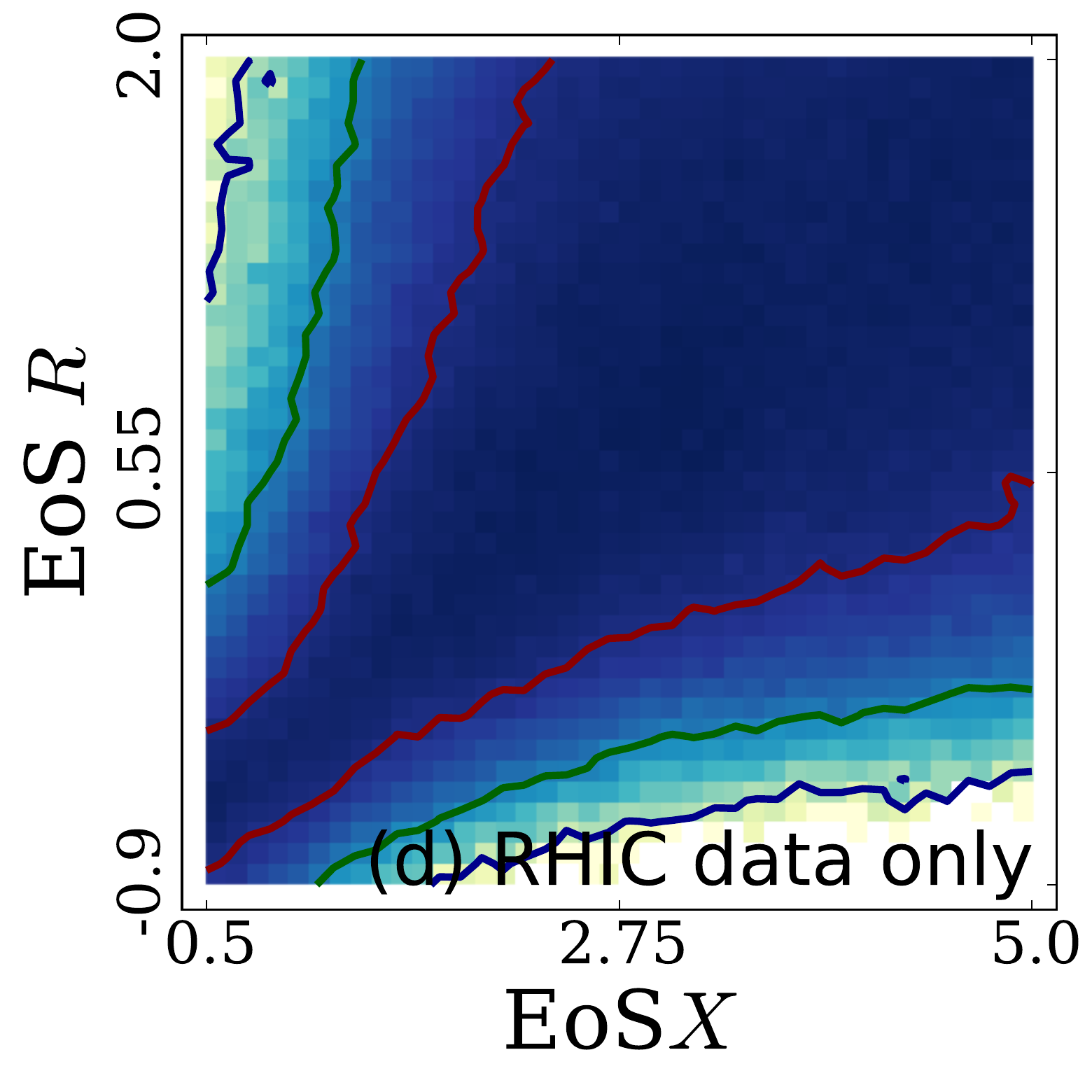}
\includegraphics[width=0.166667\textwidth]{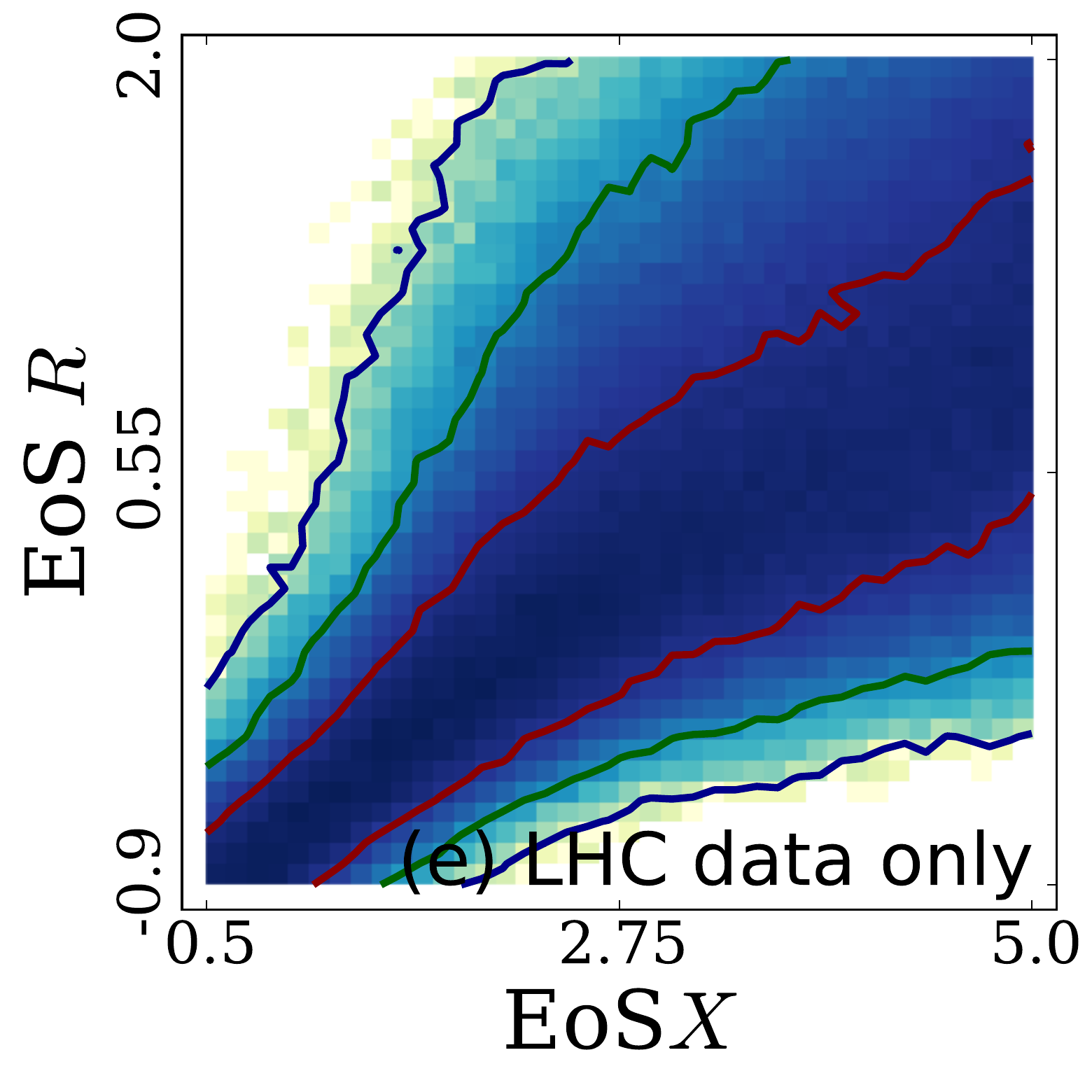}
\includegraphics[width=0.166667\textwidth]{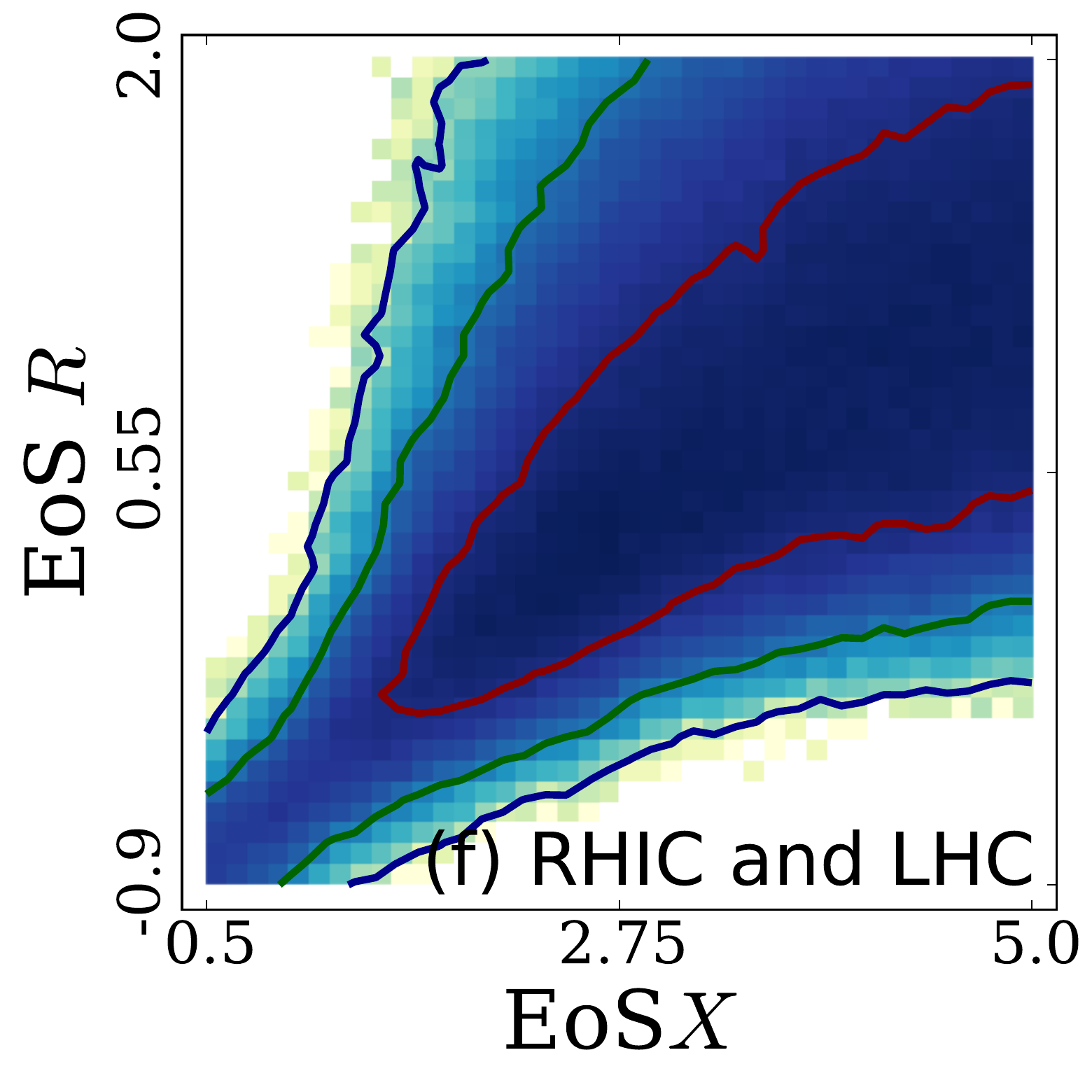}}
\caption{\label{fig:ll_RHIC_LHC}
Posterior likelihood distributions for the shear viscosity parameters generated using RHIC and LHC data separately (a \& b) and together (c). We can see that the viscosity at $T_0$ is well constrained by RHIC data alone but that there is very little constraint of its temperature dependence without the LHC data. This is consistent with the fact that LHC data probe higher temperatures. Combining the datasets clearly constrains the data better than a single data set. Panels (d-e) show corresponding information for the two equation of state parameters. In this case it is clear that LHC data provide the bulk of the resolving power. Again, this is expected because the equation of state is fixed at $T_0=165$ MeV, and LHC collisions probe regimes further into the region where the equation of state differs across the parameter space. One can also see that the high degree of covariance in both figures (c) and (f), especially for the covariance between the two equation-of-state parameters in (f). This shows that even though neither parameter is particularly well constrained, there is a linear combination that is well constrained, while the orthogonal linear combination is poorly constrained.}
\end{figure}

The principal goal of this paper is to present various methods for understanding the role certain observables, or sets of observables, play in constraining the posterior likelihood of given parameters, or sets of parameters. The most straight-forward method to determine the sensitivity is to perform the analysis both with and without a given observable, or set of observables. An example of this is illustrated in Fig. \ref{fig:ll_RHIC_LHC}. The two-dimensional posterior likelihood projection for the two viscosity parameters, $(\eta/s)_0$ and $\eta'$, are redisplayed along with projections where either RHIC data or LHC data are ignored. This makes it clear that the LHC data is especially important for constraining the temperature dependence of the viscosity, $\eta'$. 

However, the method of repeating the statistical analysis without a given observable can be onerous. If one wishes to study the sensitivity to $N_Y=30$ observables, one would repeat the MCMC procedure $N_Y$ times. Here, we present methods for finding three simple measures of the sensitivity using the output from a single MCMC trace. That information would be of the form of a list of values, $\alpha=1\cdots N_{\rm MCMC}$, where $N_{\rm MCMC}$ might be of the order of one million. For each point one would have the parametes $x_{\alpha,i}$ and the observables $y_{\alpha,a}$. 

The first measure would simply describe how a given observable, $y_a$, would change due to a change in a parameter $x_i$, while keeping all other parameters constant. Because $y_{M,a}^{\rm (\vec{x})}$ is not purely linear, one needs to choose over what region the derivative is evaluated. Two choices of interest might be the prior or the posterior. Here, we use single brackets, $\langle\cdots\rangle$, to denote an average of the prior, and double brackets, ${\langle\langle}\cdots{\rangle\rangle}$, to denote averages taken over the posterior. For averages over the prior, one can use the lists from the points at which full-model runs were performed for the purpose of training the emulator. If calculations were performed randomly throughout the prior, one could consider the covariance $\langle\delta y_a\delta x_i\rangle$, where $\delta y_a=y_a-\langle y_a\rangle$ and $\delta x_i=x_i-\langle x_i\rangle$. One can calculate the partial derivative of $y_a$ with respect to any parameter by a simple matrix inversion,
\begin{eqnarray}
\label{eq:dydx}
\langle \delta y_a\delta x_i\rangle&=&\sum_j\left\langle\frac{\partial y_a}{\partial x_j}\right\rangle\langle\delta x_j\delta x_i\rangle\\
\nonumber
\left\langle\frac{\partial y_a}{\partial x_j}\right\rangle&=&\langle \delta y_a\delta x_i\rangle \langle\delta x\delta x\rangle^{-1}_{ij}.
\end{eqnarray}
Here, the brackets around $\partial y/\partial x$ note that this is the best slope for this region of parameters space, which in this case is the prior. This expression matches the usual least-squares expression for the slope of a line in multidimensions. Equation \ref{eq:dydx} can be easily altered to address how the slope looks when focused on the posterior region by making the change $\langle\cdots\rangle\rightarrow\langle\langle\cdots\rangle\rangle$.

\subsection{Response of extracted parameter values to changes in experimental measurement}

Equation \ref{eq:dydx} shows how a specific model value, $y_a$, changes when a specific parameter, $x_i$, is varied. However, describing how a small change in an experimental measurement, $y^{\rm(exp)}_a$, affects the average value of a parameter, $\langle\langle x_i\rangle\rangle$, in the average posterior value is a different question. This involves understanding $\partial\langle\langle x_i\rangle\rangle/\partial y^{\rm(exp)}_a$. Here we consider a general expression for any posterior-averaged function of  $\vec{x}$, $\langle\langle f(\vec{x})\rangle\rangle$, which for this specific consideration will be $f(\vec{x})=x_i$.
\begin{eqnarray}
\label{eq:dxdy_general}
\frac{\partial}{\partial y^{\rm(exp)}_a}
\langle\langle f(\vec{x})\rangle\rangle
&=&\frac{\partial}{\partial y^{\rm(exp)}_a}\frac{\int d\vec{x}~f(\vec{x}){\mathcal L}(\vec{x})}{\int d\vec{x}~{\mathcal L}(\vec{x})}\\
\nonumber
&=&
\langle\langle f(\vec{x})\frac{1}{\mathcal L}\frac{\partial}{\partial y^{\rm(exp)}_a}{\mathcal L}\rangle\rangle
-\langle\langle f(\vec{x})\rangle\rangle\langle\langle \frac{1}{\mathcal L}\frac{\partial}{\partial y^{\rm(exp)}_a}{\mathcal L}\rangle\rangle.
\end{eqnarray}
Both terms on the right-hand-side of Eq. (\ref{eq:dxdy_general}) can be calculated from the trace. By setting $f=\delta x_i=(x_i-\bar{x}_i)$, with $\bar{x}_i\equiv\langle\langle x_i\rangle\rangle$, the second term vanishes in this case, and if one has an expression for the likelihood, one need only average $\delta x_i(1/{\mathcal L})\partial{\mathcal L}$ over the sampling of points from the MCMC trace to calculate the required result. One can invoke the emulator to calculate $y(\vec{x})$, and thereby determine both ${\mathcal L}$ and $\partial{\mathcal L}$. 

For Gaussian likelihoods Eq. (\ref{eq:dxdy_general}) becomes especially simple. In that case
\begin{eqnarray}
\nonumber
{\mathcal L}&=&\exp\left\{\frac{1}{2}(y_a-y_a^{\rm(exp)})\Sigma_{ab}^{-1}(y_b-y_b^{\rm(exp)})\right\}\\
\nonumber
\frac{1}{\mathcal L}\frac{\partial}{\partial y^{\rm(exp)}_a}{\mathcal L}&=&
\Sigma^{-1}_{ab}(y_a-y^{\rm(exp)}_a),\\
\label{eq:dxdy_offdiag}
\frac{\partial}{\partial y^{\rm(exp)}_a}\bar{x}_i
&=&\Sigma^{-1}_{ab}\langle\langle \delta x_i\delta y_b\rangle\rangle,\\
\nonumber
&=&\Sigma^{-1}_{ab}\left\langle\left\langle\frac{\partial y_a}{\partial x_j}\right\rangle\right\rangle A_{ji},\\
\nonumber
A_{ij}&\equiv&\langle\langle\delta x_i\delta x_j\rangle\rangle.
\end{eqnarray}
Here, $\delta y$ can be relative to any point because $\langle\langle\delta x\rangle\rangle=0$, and the derivative in the last line is the average slope for the posterior region. Eq. (\ref{eq:dxdy_offdiag}) is straightforward to calculate from the output of the MCMC. For the case where $\Sigma$ is diagonal,
\begin{equation}
\label{eq:dxdy}
\left.
\frac{\partial \bar{x}_i}{\partial y_a^{\rm(exp)}}
\right|_{y^{\rm(exp)}_{b\ne a}}=
\frac{1}{\sigma_a^2}\langle\langle\delta y_a\delta x_i\rangle\rangle
\end{equation}

\begin{figure}
\centerline{\includegraphics[width=\textwidth]{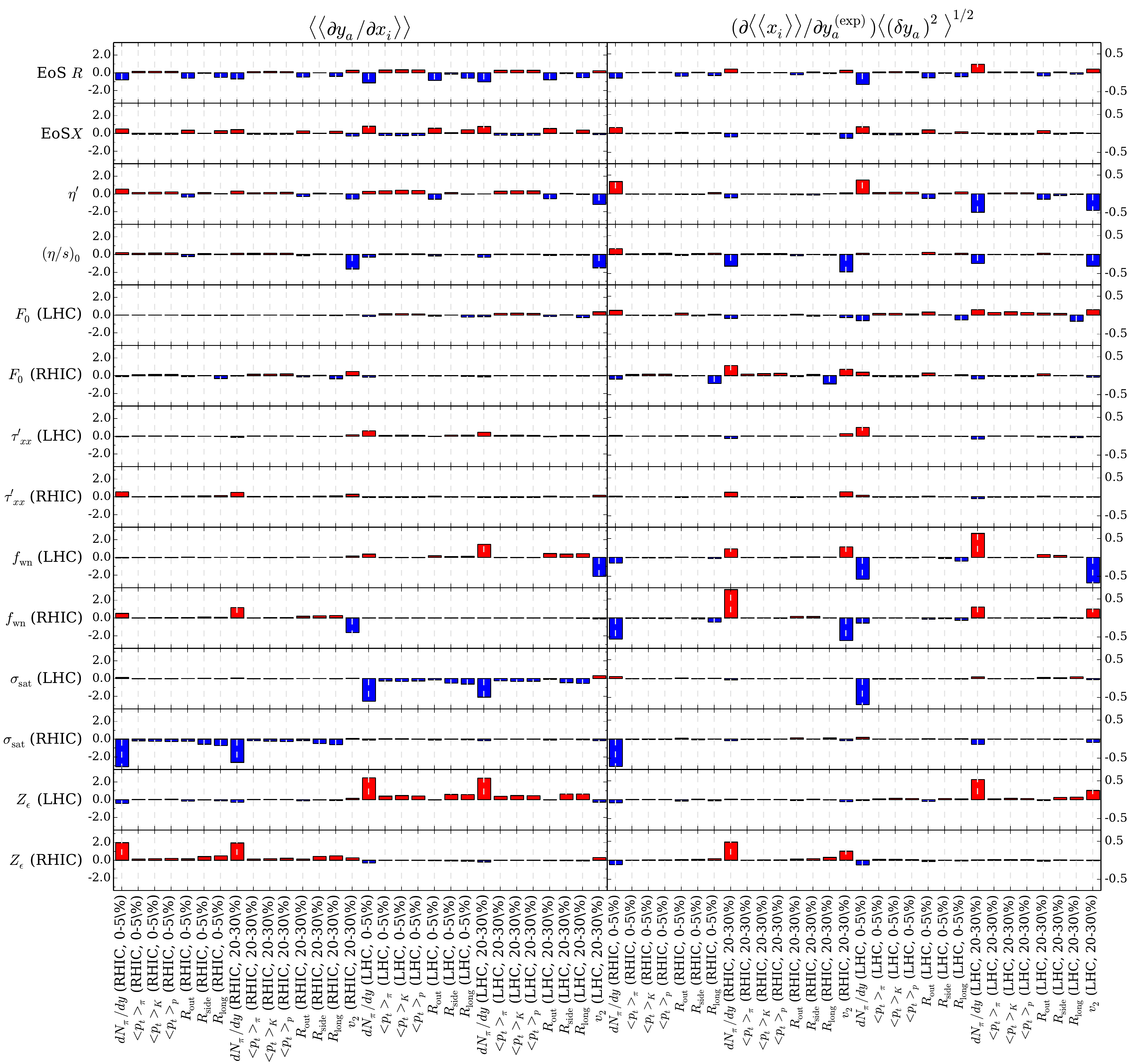}}
\caption{\label{fig:dydxdxdy}
Model responses of an observable with respect to a given parameter, $\langle\langle\partial y_a/\partial x_i\rangle\rangle$ are displayed in the left-side panel. In the right-side panel, the change of the inferred value of a parameter with respect to changes in a measurement, $\partial\langle\langle x_i\rangle\rangle/\partial y^{\rm(exp)}_a$, are scaled by the spread of model values throughout the prior, $\langle\delta y_a^2\rangle^{1/2}$. Larger absolute values point to measurements which play important roles in constraining that parameter. The signs of the response are not always equal for the corresponding derivative in the two plots.
}
\end{figure}

To better quantify how a specific observable constrains a given parameter, one may multiple the expression for $\partial \bar{x}/\partial y^{\rm(exp)}$ in Eq. (\ref{eq:dxdy}) by the measure of how much $y_a$ changes throughout the prior, $\langle\delta y_a^2\rangle^{1/2}$, because model values of an observable must change within the range of the prior if that observable is to provide resolving power. Figure \ref{fig:dydxdxdy} displays this quantity for all observables. One can also see whether the change is positive or negative, which is not obvious and may have an opposite sign compared to $\partial y/\partial x$.

As examples of the sensitivity analysis one can look at the elliptic flow measured at RHIC and at the LHC. In Fig. \ref{fig:dydxdxdy} one can see that changing the measurement of $v_2$ at RHIC strongly affects the extracted value of the viscosity at $T_0$, $(\eta/s)_0$, but has little effect on the extracted temperature dependence, $\eta'$. In contrast, the measurement of $v_2$ at the LHC more strongly affects $\eta'$, while playing a more minor role in determining $(\eta/s)_0$. Given the fact that the LHC explores higher energy densities, this was expected. One can also see that the constraints on the viscosity were driven by measurements of $v_2$ and the multiplicities, whereas constraints on the two equation-of-state parameters were driven by a wider variety of measurements. 

\subsection{Relation between experimental uncertainties and widths of posterior parameter distributions}

We now consider how the uncertainty of a specified observable affects the width of the posterior parameter distribution, i.e., calculate the derivative
\begin{eqnarray}
\label{eq:Rijab}
\frac{\partial}{\partial\Sigma_{ab}}A_{ij}
&=&\langle\langle \delta x_i\delta x_j \frac{1}{\mathcal L}\frac{\partial{\mathcal L}}{\partial\Sigma_{ab}}\rangle\rangle
-\langle\langle \delta x_i\delta x_j\rangle\rangle\langle\langle\frac{1}{\mathcal L}\frac{\partial{\mathcal L}}{\partial\Sigma_{ab}}\rangle\rangle~,
\end{eqnarray}
where the last step followed the steps from the previous subsection used to derive Eq. (\ref{eq:dxdy_general}). Equation (\ref{eq:Rijab}) is fairly straight-forward to calculate, but can be a bit cumbersome. To find a simpler, though approximate, expression we assume that the prior distribution of parameters is Gaussian  even though the results shown here all assumed hard cutoffs, and that $y(\vec{x})$ is linear, so that the posterior distribution is Gaussian. We also use Gaussian likelihoods for comparing to experimental values. When comparing calculations with and without these approximations, results changed little.

With the approximations, the overall likelihood is then of a Gaussian form,
\begin{eqnarray}
\label{eq:gaussform}
{\mathcal L}\sim \exp\left\{-|\delta x|^2/2R^2-\left\langle\left\langle\frac{\partial y_a}{\partial x_i}\right\rangle\right\rangle\Sigma_{ab}^{-1}
\left\langle\left\langle\frac{\partial y_b}{\partial x_j}\right\rangle\right\rangle
\frac{\delta x_i\delta x_j}{2}\right\},\\
\nonumber
\delta x_i=x_i-\bar{x}_i.
\end{eqnarray}
We have assumed that the parameters $x_i$ have all been scaled to have the same prior width $R$. Further, in the neighborhood of the maximum likelihood we assume that $y$ behaves linearly. The width of the posterior, $A_{ij}$, can be taken from the Gaussian form,
\begin{eqnarray}
\nonumber
A_{ij}&=&
\frac{\int d^Nx~\exp\{-A^{-1}_{ij}\delta x_i\delta x_j/2\}\delta x_i\delta x_j}
{\int d^Nx~
\exp\{-A^{-1}_{ij}\delta x_i\delta x_j/2\}},\\
\label{eq:Adef}
A_{ij}^{-1}&\equiv&\delta_{ij}/R^2+\left\langle\left\langle\frac{\partial y_a}{\partial x_i}\right\rangle\right\rangle\Sigma^{-1}_{ab}
\left\langle\left\langle\frac{\partial y_b}{\partial x_j}\right\rangle\right\rangle.
\end{eqnarray}
Our stated goal is to find an expression describing how $A_{ij}$ responds to changes in $\Sigma_{ab}$,
\begin{eqnarray}
\frac{\partial A}{\partial\Sigma_{ab}^{-1}}&=&\frac{\partial}{\partial\Sigma^{-1}_{ab}}(AA^{-1}A)\\
\nonumber
&=&2\frac{\partial}{\partial\Sigma^{-1}_{ab}}A+A\left(\frac{\partial}{\partial\Sigma_{ab}^{-1}}A^{-1}\right)A\\
\nonumber
&=&-A\left(\frac{\partial}{\partial\Sigma^{-1}_{ab}}A^{-1}\right)A\\
\label{eq:ASigmaresult}
\frac{\partial}{\partial\Sigma^{-1}_{ab}}A_{i\ell}&=&-A_{ij}\left\langle\left\langle\frac{\partial y_a}{\partial x_j}\right\rangle\right\rangle\left\langle\left\langle
\frac{\partial y_b}{\partial x_k}\right\rangle\right\rangle A_{k\ell}.
\end{eqnarray}
One can now insert the expression for $\partial y/\partial x$ given in Eq. (\ref{eq:dydx}) into Eq.s \ref{eq:Adef} and \ref{eq:ASigmaresult} to find the needed result.  When working with the hard cutoffs for priors rather than Gaussians, we scale all the parameters to have the same prior width, $\langle x^2\rangle=R^2$, in addition to being centered at zero, $\langle x\rangle=0$. If one uses the prior distribution to calculate $\partial y/\partial x$, then the response in Eq. (\ref{eq:ASigmaresult}) needn't ever access the experimental value, and one doesn't need to perform the MCMC.

\begin{figure}
\centerline{\includegraphics[width=0.6\textwidth]{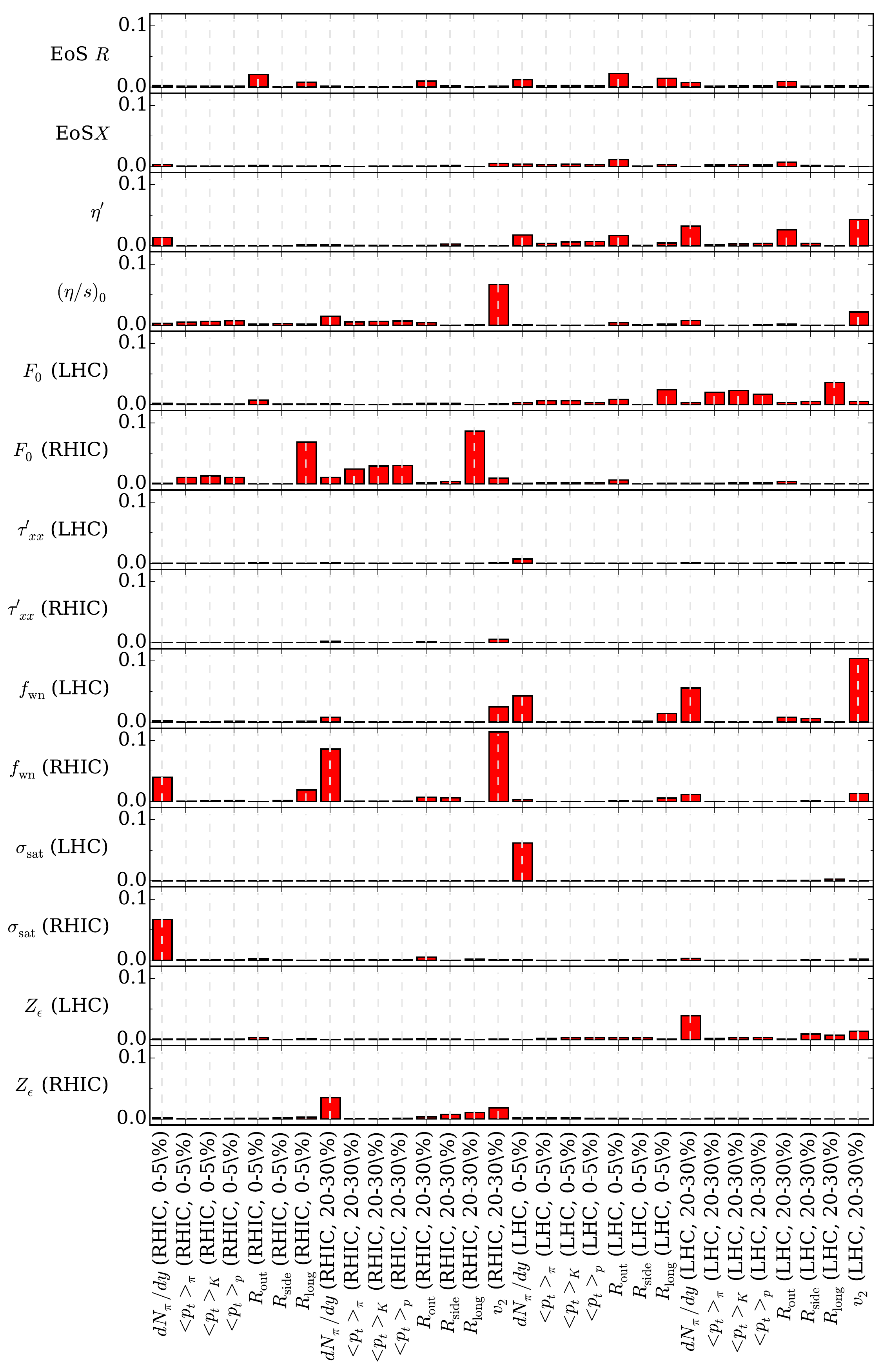}}
\caption{\label{fig:sensitivity}
The resolving power of a particular observable in determining a specific parameter, $R_{i;a}$ defined in Eq. (\ref{eq:Riadef}), is displayed for all observables and parameters.
}
\end{figure}
If one wishes to use the posterior distribution to calculate the derivatives, the form for the response simplifies
\begin{eqnarray}
\label{eq:ASigmaresult_posteriorA}
\frac{\partial}{\partial\Sigma^{-1}_{ab}}A_{ij}&=&-\langle\langle \delta x_i\delta y_a\rangle\rangle
\langle\langle\delta y_b \delta x_j\rangle\rangle,
\end{eqnarray}
which can also be transformed into an expression in terms of derivatives with respect to $\Sigma$,
\begin{eqnarray}
\nonumber
\frac{\partial}{\partial\Sigma_{ab}}A_{ij}&=&
\frac{\partial A_{ij}}{\partial\Sigma^{-1}_{cd}}\frac{\partial\Sigma^{-1}_{cd}}{\partial\Sigma_{ab}}\\
\nonumber
\frac{\partial\Sigma^{-1}_{cd}}{\partial\Sigma_{ab}}&=&\frac{\partial}{\partial\Sigma_{ab}}\left(\Sigma^{-1}_{ce}\Sigma_{ef}\Sigma^{-1}_{fd}\right)\\
\nonumber
&=&-\Sigma^{-1}_{ca}\Sigma^{-1}_{bd},\\
\label{eq:ASigmaresult_posterior}
\frac{\partial}{\partial\Sigma_{ab}}A_{ij}&=&\Sigma^{-1}_{ac}\langle\langle \delta y_c\delta x_i\rangle\rangle
\langle\langle\delta y_d \delta x_j\rangle\rangle\Sigma^{-1}_{db}.
\end{eqnarray}
In this analysis the uncertainty matrix is diagonal, $\Sigma_{ab}=\sigma_a^2\delta_{ab}$, and Eq. (\ref{eq:ASigmaresult_posterior}) can be used to calculate the resolving power for determining a single parameter $x_i$ due to a single parameter $y_a$. 
\begin{eqnarray}
\label{eq:Riadef}
R_{i;a}&\equiv&\frac{\sigma_a}{\langle(\delta x_i)^2\rangle}\frac{\partial}{\partial\sigma_a}A_{ii},\\
\nonumber
&=&\frac{\sigma_a}{\langle(\delta x_i)^2\rangle}
\Sigma^{-1}_{ac}\langle\langle \delta y_c\delta x_i\rangle\rangle
\langle\langle\delta y_d \delta x_j\rangle\rangle\Sigma^{-1}_{da}.
\end{eqnarray}
Results for $R_{i;a}$ are shown in Fig. \ref{fig:sensitivity}.

\begin{figure}
\centerline{\includegraphics[width=0.6\textwidth]{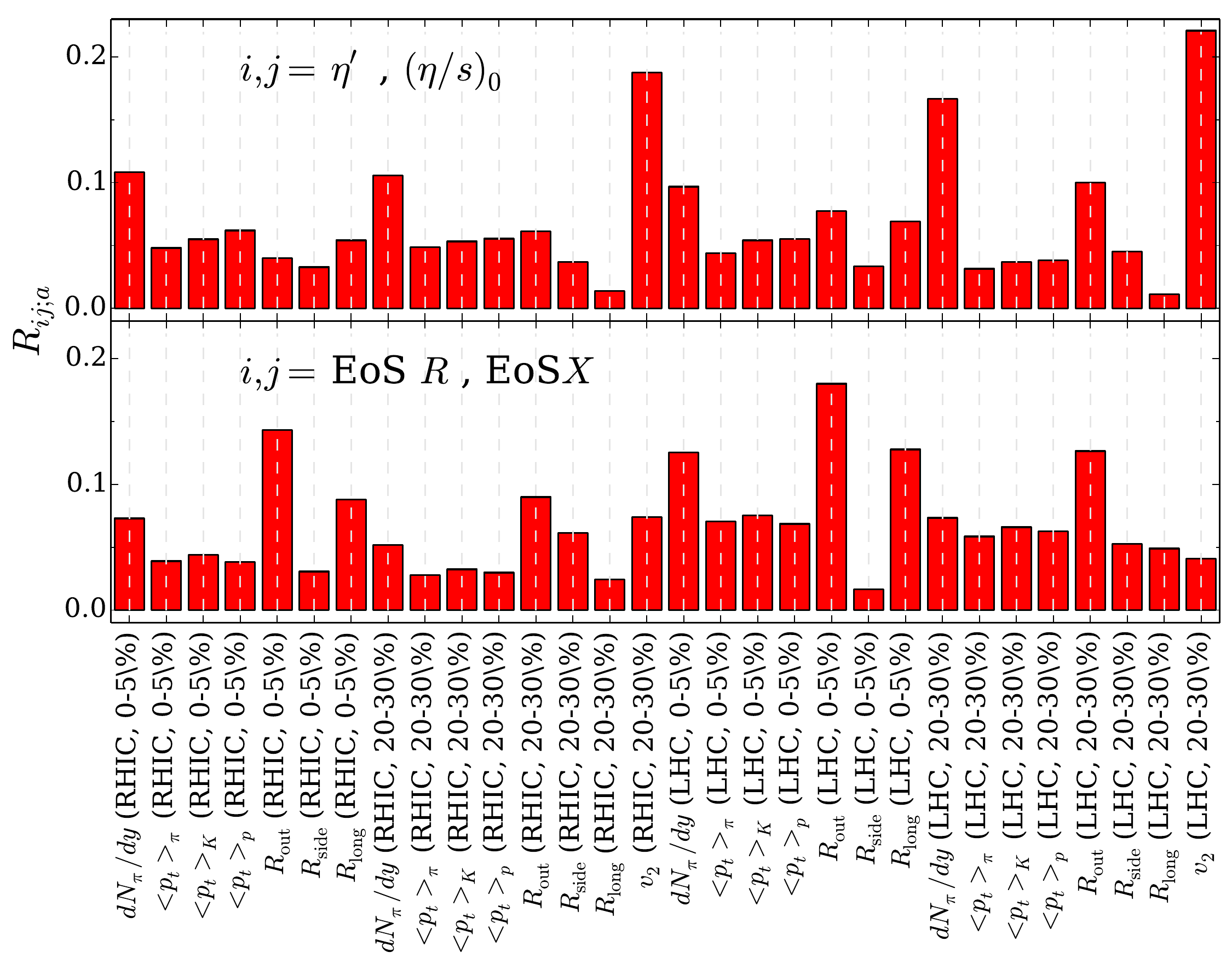}}
\caption{\label{fig:eos2eta2}
The resolving power for determining a pair of parameters, $x_i,x_j$, due to a specific observable $y_a$ as defined by $R_{ij;a}$ in Eq. (\ref{eq:Rijadef}). The sensitivity for the two viscosity parameters is shown in the upper panel and that for the two equation-of-state parameters is displayed in the lower panels. Information about eh viscosity is most strongly determined by measurements of $v_2$ and the multiplicities. The equation of state is most strongly constrained by femtoscopic radii.}
\end{figure}

Even when a specific parameter is not well constrained, a combination of that parameter with another might still be well constrained.  The two equation-of-state parameters are a good example, and the two viscosity parameters also have a strong off-diagonal component to their likelihood projections. These two instances are shown in panels (c) and (f) in Fig. \ref{fig:ll_RHIC_LHC}. If one then wants to understand the contribution of a given observable in constraining the two-dimensional projection, one needs to define a quantity which effectively measures the likely area of the two-dimensional projection onto the parameters $x_i$ and $x_j$. Here we use the determinant of the two-by-two matrix constructed from the $ij$ components of $A_{ij}$. 
\begin{eqnarray}
\label{eq:Dijdef}
D_{ij}&\equiv&A_{ii}A_{jj}-A_{ij}A_{ji},\\
\label{eq:Rijadef}
R_{ij;a}&\equiv&\sigma_a\frac{\partial}{\partial\sigma_a}D_{ij}
=\sigma_a\left\{A_{ii}\frac{\partial}{\partial\sigma_a}A_{jj}
+A_{jj}\frac{\partial}{\partial\sigma_a}A_{ii}
-2A_{ij}\frac{\partial}{\partial\sigma_a}A_{ji}
\right\}.
\end{eqnarray}
$D_{ij}$ would represent the product of the eigenvalues of the two-by-two matrix, or equivalently, the square of the area covered by the projection. The way in which $D_{ij}$ changes with respect to a given uncertainty can then be calculated with the help of Eq. (\ref{eq:ASigmaresult_posterior}). These sensitivities are illustrated in Fig. \ref{fig:eos2eta2}. This clearly demonstrates that the extracted viscosity is strongly affected by the $v_2$ measurements, and that multiplicities are also important. Other observables are of secondary, but not negligible importance. For constraining the equation of state, femtoscopic radii seem to provide the most resolving power, but all other observables contribute significantly. This underscores the importance of simultaneously considering multiple classes of observables to constrain the parameter space.

\section{Summary and Outlook}
\label{sec:summary}

The results here are unprecedented for this field, and for the first time illustrate a systematic method for identifying the critical links between parameters and observables. Remarkably, the links identified by the procedure reinforced the general knowledge of the field. For instance, it was indeed found that measurements of the elliptic anisotropy provide strong constraints of the viscosity. Further, the expectation that RHIC data would play a larger relative role in determining the viscosity near $T_c$, and that the LHC data would play a larger role in determining the temperature dependence was confirmed. The relatively large role of the multiplicities was not necessarily expected, nor was the fact that other observables provide non-negligible resolving power.

Whereas there was consensus within the field that $v_2$ would be the most important observable to determine the viscosity, the role of various observables in constraining the equation of state was very much in dispute. Figures \ref{fig:sensitivity}, \ref{fig:dydxdxdy} and \ref{fig:eos2eta2} all show that the femtoscopic source sizes provide the most resolving power. This validates ideas based on low pressures leading to more elongated phase space distributions in the outward direction and that the longitudinal sizes would increase if transverse expansion was slower, \cite{Pratt:1986cc,Pratt:2008qv,Lisa:2005dd}. Additionally, source sizes played a role in measuring the final entropy, which is also a measure of the equation of state \cite{Pal:2003rz}. The mean transverse momentum had long ago been pointed out as being sensitive to the temperature, and therefore the equations of state, \cite{VanHove:1982vk}, and even $v_2$ had been suggested as a means for extracting early pressure \cite{Sorge:1996pc}. Therefore, it was not surprising to see the resolving power for the equation of state in Fig. \ref{fig:eos2eta2} spread across all observables. 

The techniques presented here could play a pivotal role in determining the direction of future experiments. Before embarking on an expensive experimental program to improve the measure of a specific observable, one could check to see how that improvement might indeed better determine parameters of greatest interest. For example, one could determine whether running an accelerator for a specific projectile target combination, or at a new beam energy, or with higher statistics would best provide insight into better determining the equation of state. In many cases it would behoove the community to pre-analyze a project with modern statistical techniques before investing the cost and manpower for the effort.



\begin{acknowledgments}
This work was supported by the National Science Foundation's Cyber-Enabled Discovery and Innovation Program through grant NSF-0941373 and by the Department of Energy Office of Science through grant number DE-FG02-03ER41259.
\end{acknowledgments}

\end{document}